\pdfoutput=1
\documentclass[prd,10pt,nofootinbib,twocolumn,superscriptaddress,preprintnumbers,balancelastpage,longbibliography]{revtex4-1}

\usepackage{amsmath,amssymb}	
\usepackage{mathtools}
\usepackage{slashed}
\usepackage{xspace}
\usepackage{comment}
\usepackage{braket}
\usepackage{graphicx}
\usepackage{float}
\usepackage{siunitx}
\usepackage{physics}
\usepackage{fontawesome}
\usepackage{booktabs}
\usepackage{tikz}
\usetikzlibrary{calc}
\usepackage{anyfontsize}
\usepackage{aas_macros}

\definecolor{linkcolor}{rgb}{0.0, 0.28, 0.67}

\usepackage[
   colorlinks=true,
    urlcolor=linkcolor,
   anchorcolor=linkcolor,
    citecolor=linkcolor,
    filecolor=linkcolor,
    linkcolor=linkcolor,
    menucolor=linkcolor,
    linktocpage=true,
    pdfproducer=medialab,
    pdfa=true
]{hyperref}

\DeclareSIUnit\electronvolt{e\kern-.05em V}
\definecolor{modelParamsColor}{HTML}{565adb}
\definecolor{CMBarrowColor}{HTML}{c71c22}

\defcitealias{Giovanetti_2024}{Paper~I}

\begin{document}

\title{Cosmological Parameter Estimation with a Joint-Likelihood Analysis of the \\ Cosmic Microwave Background and Big Bang Nucleosynthesis}
\author{Cara Giovanetti}
\email{cg3566@nyu.edu}
\thanks{ORCID: \href{https://orcid.org/0000-0003-1611-3379}{0000-0003-1611-3379}}
\affiliation{Center for Cosmology and Particle Physics, Department of Physics, New York University, New York, NY 10003, USA}

\author{Mariangela Lisanti}
\email{mlisanti@princeton.edu}
\thanks{ORCID: \href{https://orcid.org/0000-0002-8495-8659}{0000-0002-8495-8659}}
\affiliation{Department of Physics, Princeton University, Princeton, NJ 08544, USA}
\affiliation{Center for Computational Astrophysics, Flatiron Institute, 162 Fifth Ave, New York, NY 10010, USA}

\author{Hongwan Liu}
\email{hongwan@bu.edu}
\thanks{ORCID: \href{https://orcid.org/0000-0003-2486-0681}{0000-0003-2486-0681}}
\affiliation{Physics Department, Boston University, Boston, MA 02215, USA}
\affiliation{Kavli Institute for Cosmological Physics, University of Chicago, Chicago, IL 60637}
\affiliation{Theoretical Physics Department, Fermi National Accelerator Laboratory, Batavia, IL 60510}
\author{Siddharth Mishra-Sharma}
\email{smsharma@mit.edu}
\thanks{ORCID: \href{https://orcid.org/0000-0001-9088-7845}{0000-0001-9088-7845}}
\thanks{Currently at Anthropic; work performed while at MIT/IAIFI.}
\affiliation{The NSF AI Institute for Artificial Intelligence and Fundamental Interactions}
\affiliation{Center for Theoretical Physics, Massachusetts Institute of Technology, Cambridge, MA 02139, USA}
\affiliation{Department of Physics, Harvard University, Cambridge, MA 02138, USA}

\author{Joshua T. Ruderman}
\email{ruderman@nyu.edu}
\thanks{ORCID: \href{https://orcid.org/0000-0001-6051-9216}{0000-0001-6051-9216}}
\affiliation{Center for Cosmology and Particle Physics, Department of Physics, New York University, New York, NY 10003, USA}

\preprint{MIT-CTP/5737}

\date{\today}

\begin{abstract} 
We present a joint-likelihood analysis of Big Bang Nucleosynthesis (BBN) and Cosmic Microwave Background (CMB) data, consistently combining likelihoods and taking into account uncertainties in nuclear reaction rates for the first time.  Bayesian inference is performed on the baryon abundance and the effective number of neutrino species, $N_{\rm eff}$, using a CMB Boltzmann solver in combination with LINX, a new flexible and efficient BBN  code. We marginalize over Planck nuisance parameters and nuclear rates to find $N_{\rm{eff}} = 3.08_{-0.14}^{+0.15},\,2.94 _{-0.15}^{+0.16},$ or $2.96_{-0.14}^{+0.13}$, for three separate reaction networks.  This framework enables robust testing of the Lambda Cold Dark Matter paradigm and its variants with CMB and BBN data.
\end{abstract}
\maketitle

\noindent
\noindent\textbf{Introduction.---}Both the Cosmic Microwave Background~(CMB) and the primordial abundance of light elements produced during Big Bang Nucleosynthesis~(BBN) provide a wealth of information about the early Universe and its constituents.  Precision measurements of the CMB have tightly constrained the Lambda Cold Dark Matter~($\Lambda$CDM) cosmological parameters, with corroborating measurements from BBN painting a largely consistent picture of the Universe at early times. 

The CMB power spectrum is at the forefront of precision cosmology, providing percent-level measurements of the  $\Lambda$CDM cosmological parameters (\textit{e.g.}\ the baryon abundance, $\Omega_b h^2$) and  constraining alternative scenarios (\textit{e.g.}\ changes to the effective number of neutrino species, $N_\text{eff}$).  Analyses performed on Planck data~\cite{Planck2018} often employ Bayesian inference using sampling methods like Markov-Chain Monte Carlo~(MCMC) or nested sampling to explore the high-dimensional parameter space, including a large number of nuisance parameters. Such detail is now expected even in studies exploring new-physics signatures, such as  Refs.~\cite{Alvi:2022aam, Bansal:2021dfh}.

Meanwhile, BBN provides a unique window into the epoch when the baryon temperature was in the \SI{}{\kilo\eV}--\SI{}{\mega\eV} range. 
The baryon density can impact primordial element abundances, since a higher density allows newly-formed nuclei to interact more frequently, fusing them into heavier elements.
Recent experimental determinations of D/H (the  number density ratio of deuterium to hydrogen nuclei)~\cite{Cooke_2016,Riemer_S_rensen_2017,Cooke_2018,Zavarygin_2018} and Y$_\mathrm{P}$ (the mass abundance of helium-4)~\cite{Aver_2015,Aver_2021,Valerdi_2019,Fern_ndez_2019,Kurichin_2021,Hsyu_2020,Valerdi_2021}) have reached percent-level precision. 
This, combined with theoretical predictions for the abundance of these nuclides using public BBN codes~\cite{Arbey_2012,Pitrou_2018,Consiglio_2018,arbey2019,burns_2023}, allows for a percent-level determination of $\Omega_b h^2$,  rivaling that of the CMB\@. 
Uncertainties in the theoretical predictions are dominated by the uncertainty in the rates of nuclear reactions (including the neutron lifetime) occurring during BBN; these have been studied carefully for many different reactions in \textit{e.g.}~Refs.~\cite{GomezInesta_2017,deSouza_2019,deSouza_2019b,Rijal_2019,deSouza_2020,Mossa_2020,Pisanti_2021,Moscoso_2021}. 
In practice, each rate's uncertainty is modeled by a nuisance parameter; accurate Y$_\mathrm{P}$ and D/H predictions require marginalizing over parameters for all reactions that significantly influence these abundances. 
BBN can also provide an independent constraint on $N_{\rm{eff}}$, since a larger $N_\mathrm{eff}$ increases the expansion rate, increasing Y$_\mathrm{P}$ by reducing the time available for weak interactions to convert neutrons to protons, and increasing D/H by ending deuterium burning earlier.

Previous work has extensively demonstrated the importance of BBN to the $\Lambda$CDM and $\Lambda$CDM$\texttt{+}N_\mathrm{eff}$ models.  The Planck Collaboration performed a joint CMB power spectrum and BBN analysis, in which they incorporate information from both epochs to perform parameter estimation.  This analysis marginalized over all Planck experimental nuisance parameters, but only estimated the effects of nuclear rate uncertainties using either a constant theory uncertainty on the D/H and Y$_\text{P}$ predictions~\cite{Planck2018} or marginalizing over just one reaction rate~\cite{Planck2015}. On the other hand, BBN-focused analyses (\textit{e.g.} Refs.~\cite{Nollett_2011,Pitrou_2018,Pitrou_2021,Pisanti_2021}) better capture these uncertainties in the predicted BBN abundances by varying these nuclear rates during parameter estimation.  This latter class of analyses often uses only a CMB prior on $\Omega_b h^2$, and so Ref.~\cite{Yeh_2022} advances this class of analysis a step further by also including CMB priors on $N_{\rm{eff}}$ and Y$_\textrm{P}$.

However, none of these analyses can account for correlated uncertainties between cosmological parameters, CMB experimental nuisance parameters, and BBN nuisance parameters, which can only be captured by varying all model and nuisance parameters in both epochs.  Such an analysis has been prohibitively expensive in the past because publicly-available BBN codes either have relatively long runtimes or are structured such that nuclear rate uncertainties cannot be easily varied, making them difficult to combine with standard parameter estimation frameworks such as MCMC samplers.

Furthermore, the CMB power spectrum is dependent on Y$_\mathrm{P}$ through its impact on the damping tail. Therefore, correctly predicting the power spectrum for given values of $\Omega_b h^2$ and $N_\mathrm{eff}$ requires the BBN prediction of Y$_\mathrm{P}$ at these parameter values. This is not accounted for in BBN-focused analyses. For CMB-focused analyses, Boltzmann codes such as CLASS~\cite{lesgourgues_2011a,lesgourgues_2011b,lesgourgues_2011c,lesgourgues_2011d} or CAMB~\cite{Lewis_1999,Howlett_2012} rely on a table of predicted Y$_\mathrm{P}$ as a function of $\Omega_b h^2$ and $N_\mathrm{eff}$.  This does not properly capture the uncertainty in the BBN prediction, and also must be recomputed for other alternatives to $\Lambda$CDM\@.   

Ref.~\cite{Giovanetti_2024} (hereinafter \citetalias{Giovanetti_2024}) introduces the new public BBN code LINX~(Light Isotope Nucleosynethesis with JAX) and its underlying computational formalism. 
LINX is written with JAX \cite{jax2018github,deepmind2020jax}---a Python-based framework that allows for fast compiled code---and is structured for efficient parameter estimation using MCMC samplers.  
This work uses BBN abundance predictions from LINX to perform Bayesian inference on cosmological parameters. 
We perform a BBN-only inference of $\Omega_b h^2$ (with $N_\mathrm{eff}$ fixed) and $\Omega_b h^2 \texttt{+} N_{\rm{eff}}$, marginalizing over the nuclear rate uncertainties.  
We then generalize these to CMB\texttt{+}BBN analyses, using the Boltzmann code CLASS \cite{lesgourgues_2011a,lesgourgues_2011b,lesgourgues_2011c,lesgourgues_2011d} for the CMB power spectrum prediction. 
The joint Bayesian analysis---the first of its kind---varies all cosmological parameters, Planck nuisance parameters, and BBN nuisance parameters, while including the BBN prediction of Y$_{\rm{P}}$ in the CMB power spectrum prediction.  

\noindent
\noindent\textbf{BBN-only Analysis.---}   
To begin, we perform a BBN-only study to infer $\Omega_bh^2$ (with $N_\text{eff}$ fixed) and $\Omega_bh^2 \texttt{+} N_{\rm{eff}}$ using LINX\@. In the latter, we add an inert, relativistic species that can have a positive or negative energy density (negative energy density being an unphysical scenario that can, however, mimic colder-than-expected neutrinos), following \textit{e.g.} Refs.~\cite{Planck2018,Yeh_2022}.  The rate for each nuclear reaction $i$ is $r_i(T) \equiv u^{-1} \langle \sigma v \rangle (T)$, where $u$ is the atomic mass unit, $\langle \sigma v \rangle$ is the velocity-averaged cross section, and $T$ is the baryon temperature---see \textit{e.g.}\ Ref.~\cite{Pitrou_2018} for discussion of the unit convention. 
This work adopts three separate sets of reaction rates: \textit{1)}~``PRIMAT'', the default small network used by the PRIMAT BBN code~\cite{Pitrou_2018}, \textit{2)}~``PArthENoPE'', a set of rates extracted from the PArthENoPE BBN code~\cite{Pisanti_2021}, and \textit{3)}~``YOF'', a set of rates used by Ref.~\cite{Yeh:2020mgl}, provided by the PRyMordial BBN code~\cite{burns_2023}. 
Following  Refs.~\cite{Pitrou_2018,Fields:2019pfx}, the rate uncertainties are captured by taking $r_i$ to be log-normally distributed, with a mean value $ \overline{r}_i(T)$ and standard deviation $\sigma_i(T)$. 
Specifically, $\log r_i (T) = \log \overline{r}_i (T) + q_i \sigma_i (T)$, where $q_i$ is a unit Gaussian random variable.  The BBN nuisance parameters, denoted as  $\boldsymbol{\nu}_\text{BBN}$, are $q_i$ and the neutron  lifetime. 

To obtain $\overline{r}_i(T)$ and $\sigma_i(T)$, measurements of the cross section as a function of energy are combined across different experiments, and a fit to the data is performed before velocity-averaging. 
Each set of rates makes different choices on the fitting procedure and the experimental data that are used.  
PArthENoPE performs purely phenomenological polynomial fits to the data~\cite{Pisanti_2021}. 
PRIMAT fits the data using functional forms derived from \textit{ab initio} theory calculations~\cite{GomezInesta_2017,Moscoso_2021}.  YOF relies mainly on results provided by the NACREII Collaboration, which uses experimental data to fit a potential model for each nuclear reaction~\cite{Xu_2013}.
All three sets incorporate the recent measurement of the $d(p, \gamma)^3\mathrm{He}$ rate from \SIrange{30}{300}{\kilo\eV} by the LUNA Collaboration~\cite{Mossa_2020}.

Because of the different approaches taken, these sets of rates also make different predictions. 
D/H predictions based on PRIMAT are in $1.8\sigma$ tension with the observed  abundance, whereas predictions based on PArthENoPE and YOF 
are not~\cite{Pisanti_2021,Moscoso_2021}.  
Experimental determinations of $d(d,n)^3$He and $d(d,p)t$ will be crucial for reducing the uncertainty in the prediction of D/H, and 
may resolve these discrepancies~\cite{Pitrou:2021vqr,Moscoso_2021,Pisanti_2021}.

We construct a BBN likelihood using the observed D/H determined by Ref.~\cite{Cooke_2018} and Y$_{\rm{P}}$ from Ref.~\cite{Aver_2015}:
\begin{alignat*}{2}
    \textrm{D/H}^{\rm{obs}} &= 2.527\times10^{-5}\hspace{0.5cm}
    \sigma_{\textrm{D/H}^{\rm{obs}}} &&= 0.030\times10^{-5}\\
    \textrm{Y}_{\textrm{P}}^{\rm{obs}} &=0.2449  \hspace{2cm}
    \sigma_{\textrm{Y}_{\textrm{P}}^{\rm{obs}}} &&=0.004 \,.
\end{alignat*}
The BBN log-likelihood is given by
\begin{multline}
    -2 \log\mathcal{L}_{\rm{BBN}}= \left(\frac{\textrm{Y}_{\textrm{P}}^{\rm{pred}}(\Omega_b h^2,N_{\rm{eff}},\boldsymbol{\nu}_{\rm{BBN}})-\textrm{Y}_{\textrm{P}}^{\rm{obs}}}{\sigma_{\textrm{Y}_{\textrm{P}}^{\rm{obs}}}}\right)^2 \\+ 
    \left(\frac{\textrm{D/H}^{\rm{pred}}(\Omega_b h^2,N_{\rm{eff}},\boldsymbol{\nu}_{\rm{BBN}})-\textrm{D/H}^{\rm{obs}}}{\sigma_{\textrm{D/H}^{\rm{obs}}}}\right)^2 \,,\label{eq:BBN_like}
\end{multline}
where LINX computes $\textrm{Y}_{\textrm{P}}^{\rm{pred}}(\Omega_b h^2,N_{\rm{eff}},\boldsymbol{\nu}_{\rm{BBN}})$ and $\textrm{D/H}^{\rm{pred}}(\Omega_b h^2,N_{\rm{eff}},\boldsymbol{\nu}_{\rm{BBN}})$.

We use dynesty~\cite{Speagle_2020,speagle_2023} to perform nested sampling~\cite{Skilling_2004,Skilling_2006}, specifying bounding and static sampling methods 
as in Refs.~\cite{Feroz_2009,Neal_2003,Handley_2015a,Handley_2015b}.
Nested sampling for parameter inference has a number of advantages over methods like MCMC, chief among them adaptability for complex, multimodal posteriors and ease of computing Bayesian evidence~(see \textit{e.g.}~\cite{Ashton_2022}).  
This facilitates marginalization over the nuisance parameters in our analyses.  Details of the computational analysis are discussed in Appendix~\ref{app:details}\@. 

We include 12 key nuclear reactions to determine $\textrm{Y}_{\textrm{P}}^{\rm{pred}}(\Omega_b h^2,N_{\rm{eff}},\boldsymbol{\nu}_{\rm{BBN}})$ and $\textrm{D/H}^{\rm{pred}}(\Omega_b h^2,N_{\rm{eff}},\boldsymbol{\nu}_{\rm{BBN}})$; including additional reactions results in changes to the predictions that are well within experimental uncertainties (see~\citetalias{Giovanetti_2024}).  
Including the neutron lifetime and the cosmological parameters $\Omega_b h^2$~(and $N_\mathrm{eff}$), there are 14~(15) parameters in total.  
We choose flat priors for  cosmological parameters, a Gaussian prior of $879.4\pm\SI{0.6}{\second}$~\cite{Zyla2020} for the neutron lifetime, and a unit Gaussian prior for all 12 $q_i$'s.

\begin{figure}[t]
    \centering
    \includegraphics[width=\linewidth]{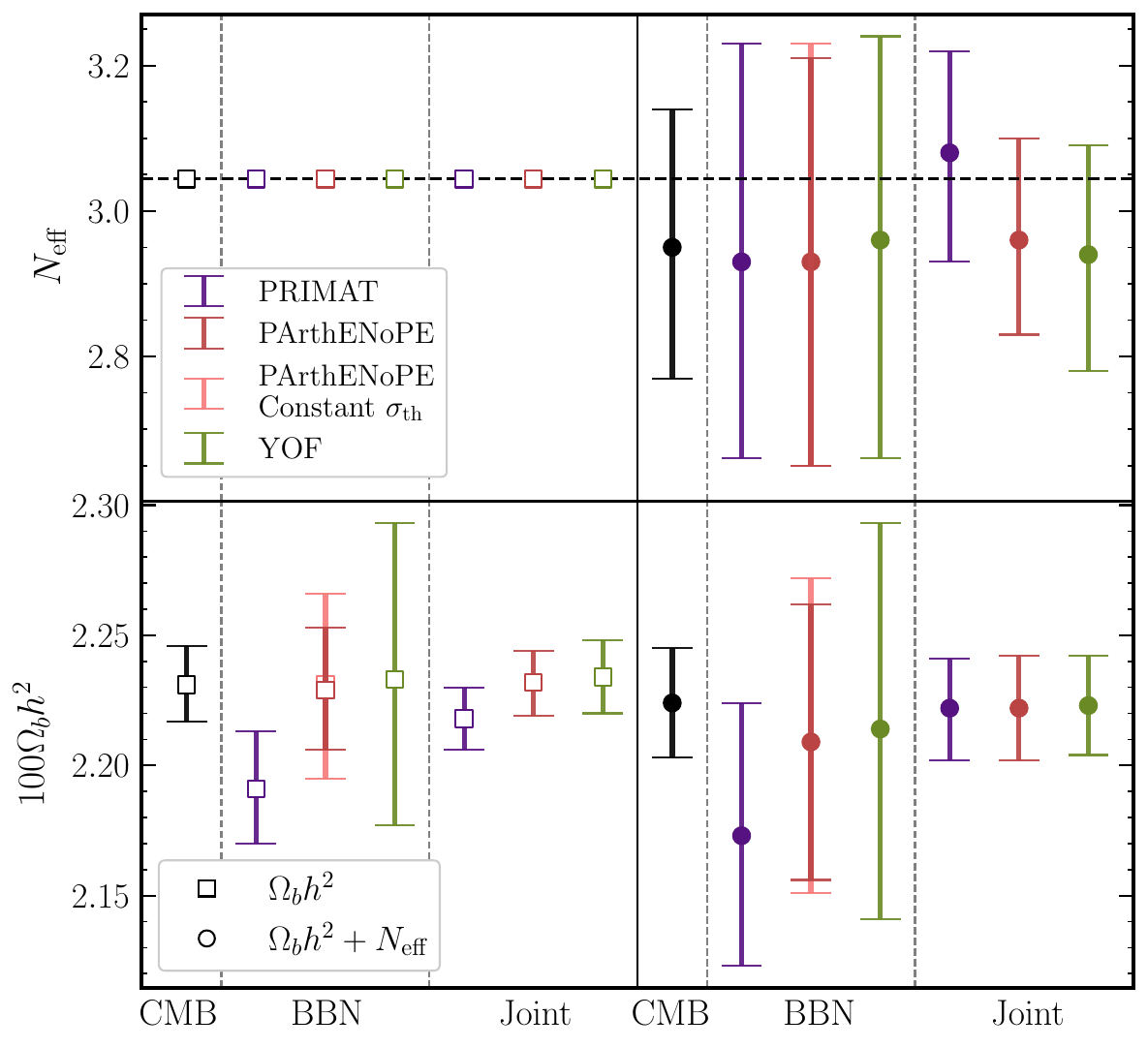}
    \caption{Parameter medians and 68\% credible limits with the inclusion of Planck data only, BBN data only, and both BBN and Planck data, as labeled at the bottom of the figure. Values from individual analyses line up top-to-bottom. Square markers in the left set of columns indicate $N_{\rm{eff}}$ is held fixed, while circular markers in the right set of columns indicate analyses where $N_{\rm{eff}}$ is allowed to float. The $\Lambda$CDM value of $N_{\rm{eff}}=3.045$ is indicated with the black dashed line.  Purple, red, and green points use the PRIMAT, PArthENoPE, and YOF networks, respectively. Pink points  indicate that a constant uncertainty is added to the deuterium part of the BBN likelihood. Otherwise, all available nuisance parameters are marginalized over. This figure summarizes the parameter values from analyses in this work and their sensitivity to different data sets and BBN reaction networks; illustrating \textit{e.g.} shrinking error bars relative to BBN when CMB data is included and systematics from using different BBN reaction networks.}
    \label{fig:parameter_evolution}
\end{figure}

Fig.~\ref{fig:parameter_evolution} summarizes the main results of this paper; numerical values for these and other analyses are compiled in Appendix~\ref{app:results} for reference. To validate LINX, we compare the results in the ``BBN only'' column with those reported in Ref.~\cite{schoneberg2024}; for the PRIMAT and YOF networks, there is excellent agreement for both $\Omega_b h^2$ and $\Omega_b h^2 \texttt{+} N_\text{eff}$. 

\begin{figure}[t]
    \centering
    \begin{tikzpicture}

    \coordinate (modelParams) at (-5,0);
    \node[anchor=west] (params) at (modelParams) {$\begin{pmatrix} \textcolor{modelParamsColor}{\Omega_b h^2} \\ \textcolor{modelParamsColor}{N_{\rm{eff}}}\end{pmatrix}$};
    \node[align=center][anchor=south,yshift=.5cm,scale=0.9] at (params) {Common\\model\\parameters};
    \coordinate (arrowStart) at ($(params.east)$);

    \draw[-,thick,modelParamsColor] (arrowStart) -- ++(0.5,0) coordinate (littleLine);

    \draw[-,thick,modelParamsColor] (littleLine) -- ++(0,-1.2) coordinate (ArrowIntermediate);
    \draw[->,thick,modelParamsColor] (ArrowIntermediate) -- ++(0.5,0) node[align=center, anchor=north, xshift=-0.25cm,yshift=-0.1cm,scale=0.9] {CMB\\input} coordinate (endOfCMBModelInputArrow);
    \draw[-,thick,modelParamsColor] (littleLine) -- ++(0,1.2) coordinate (ArrowIntermediate2);
    \draw[->,thick,modelParamsColor] (ArrowIntermediate2) -- ++(0.5,0) node[align=center, anchor=south, xshift=-0.25cm,yshift=.04cm,scale=0.9] {BBN\\input} coordinate (endOfBBNModelInputArrow);

    \node[anchor=west] (BBNparams) at (endOfBBNModelInputArrow) {$\begin{pmatrix} \textcolor{modelParamsColor}{\Omega_b h^2} \\ \textcolor{modelParamsColor}{N_{\rm{eff}}} \\ \boldsymbol{\nu}_{\rm{BBN}}\end{pmatrix}$};
    \coordinate (BBNinputarrowStart) at ($(BBNparams.east)$);

    \node[anchor=west] (CMBparams) at (endOfCMBModelInputArrow) {$\begin{pmatrix} \textcolor{modelParamsColor}{\Omega_b h^2} \\ \textcolor{modelParamsColor}{N_{\rm{eff}}} \\ \textcolor{CMBarrowColor}{\rm{Y}_{\rm{P}}} \\ \Omega_c h^2 \\ H_0 \\ A_s \\ n_s \\ \tau \\ \boldsymbol{\nu}_{\rm{CMB}}\end{pmatrix}$};
    \coordinate (CMBinputarrowStart) at ($(CMBparams.east)$);
    \coordinate (CMBYp) at ($(CMBparams.east)+(-.5,.57)$);

    \draw[->,thick] (BBNinputarrowStart) -- ++(0.7,0) node[midway, above, sloped, align=center,scale=0.9,yshift=0.1cm] {BBN\\solver} coordinate (endOfBBNInputArrow);

    \draw[->,thick] (CMBinputarrowStart) -- ++(2.8,0) node[midway, below, sloped, align=center,scale=0.9] {Boltzmann solver} coordinate (endOfCMBInputArrow);

    \node[anchor=west] (BBNoutput) at (endOfBBNInputArrow) {$\begin{pmatrix}  \rm{D/H}^{\rm{pred}}\\
    \textcolor{CMBarrowColor}{\rm{Y}_{\rm{P}}^{\rm{pred}}} \end{pmatrix}$};
    \coordinate (BBNoutputarrowStart) at ($(BBNoutput.east)+(0,0)$);
    \coordinate (YparrowStart) at ($(BBNoutput.south)+(-0.02,-0.01)$);

    \node[anchor=west] (CMBoutput) at (endOfCMBInputArrow) {$\begin{pmatrix} \vec{C}_{\ell\rm{,pred}}^{\rm{TT}} \\ \vec{C}_{\ell\rm{,pred}}^{\rm{TE}}\\
    \vec{C}_{\ell\rm{,pred}}^{\rm{EE}}
    \end{pmatrix}$};
    \coordinate (CMBoutputarrowStart) at (CMBoutput.north);

    \draw[-,thick,CMBarrowColor] (YparrowStart) -- ++(0,-1.14) coordinate (ArrowIntermediate3);
    \draw[->,thick,CMBarrowColor] (ArrowIntermediate3) -- node[midway, above, sloped, align=center,scale=0.9] {CMB\\input} ++(-1.65,0);

    \draw[-,thick] (BBNoutputarrowStart) -- ++(.35,0) coordinate (ArrowIntermediate4);
    \draw[->,thick] (ArrowIntermediate4) -- ++(0,-.55) coordinate (endOfBBNOutputArrow);
    \node[anchor=west,yshift=-0.12cm,xshift=0.01cm,align=left,scale=0.9] at (ArrowIntermediate4) {Joint\\likelihood};

    \draw[->,thick] (CMBoutputarrowStart) -- ++(0,.55) coordinate (endOfCMBOutputArrow);

    \node[anchor=north,xshift=0.5cm] at (endOfBBNOutputArrow) {$\mathcal{L}_{\rm{BBN}}\cdot\mathcal{L}_{\rm{CMB}}$};

    \end{tikzpicture}
    \caption{A schematic illustrating the joint likelihood used for the CMB\texttt{+}BBN analyses.  Note that each likelihood is computed at the same values of the input parameters $\Omega_b h^2$ and $N_{\rm{eff}}$, and the Y$_\mathrm{P}^\mathrm{pred}$ output of the BBN solver is used as input to the Boltzmann solver to ensure consistency.}
    \label{fig:likelihood}
\end{figure}
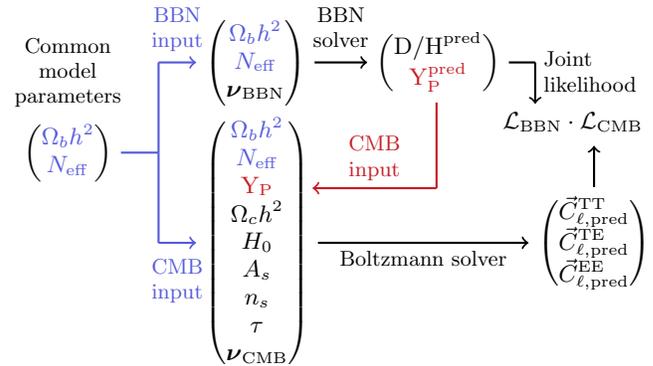

When using the PRIMAT and YOF networks, Ref.~\cite{schoneberg2024} marginalizes over the nuclear rate uncertainties.  However, because of the structure of the PArthENoPE code, Ref.~\cite{schoneberg2024}, along with Refs.~\cite{Planck2018,Pisanti_2021}, uses a constant theory uncertainty $\sigma_{\rm th}$ on D/H$^{\rm{pred}}$ to estimate the nuclear rate uncertainties on the PArthENoPE rates.  Here, to mimic this simplified procedure, the theory uncertainty $\sigma_\mathrm{th}$ is computed by sampling the relevant rates at a particular parameter combination that is expected to be near the mean of the posterior~\cite{Planck2018,Pisanti_2021,schoneberg2024}.  

\begin{figure}[t]
    \centering
    \includegraphics[width=\linewidth]{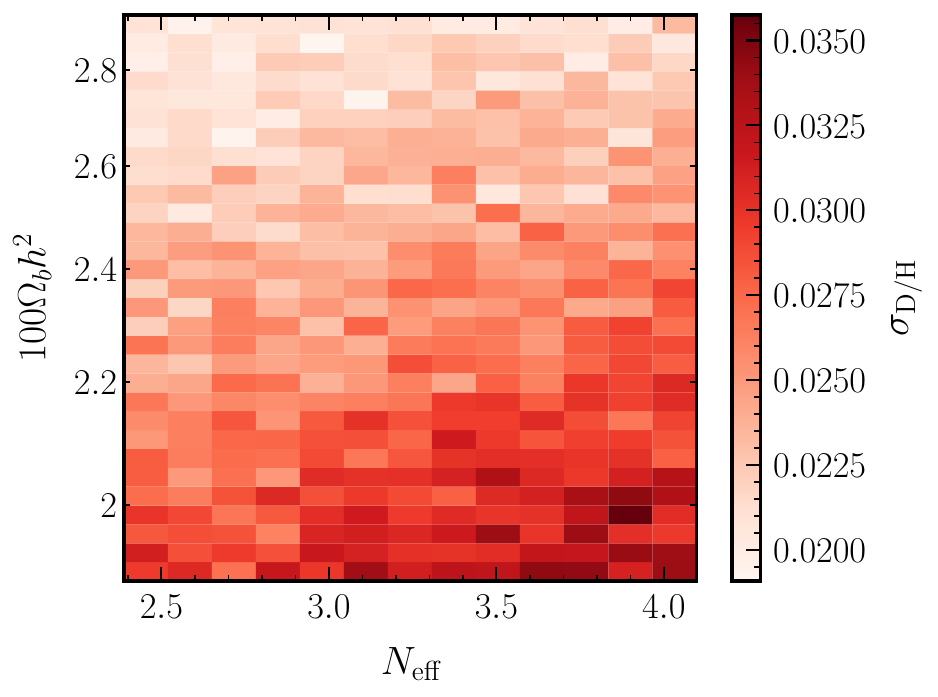}
    \caption{Variation in the standard deviation for the prediction of D/H in the PRIMAT network as a function of model parameters $N_{\rm{eff}}$ and $\Omega_b h^2$, obtained by sampling 200 different sets of values for the BBN nuisance parameters at each point in the grid.}
    \label{fig:sigma_ydp}
\end{figure}

However, the uncertainty in the D/H prediction is not constant with respect to $\Omega_bh^2$ and $N_{\rm{eff}}$.  Fig.~\ref{fig:sigma_ydp} depicts the uncertainty on D/H in the PRIMAT network, calculated using LINX\@.  This data is obtained by sampling 200 different sets of values for the nuisance parameters at each point on the grid and then finding the standard deviation $\sigma_\text{D/H}$ of the resulting D/H at each parameter point.  We find $\sigma_\text{D/H}$ fluctuates by $\sim 30\%$ throughout this range; this is in contrast to, for example, the adoption of a constant $\sigma_{\rm{D/H}}\simeq 0.03$ in Ref.~\cite{Planck2018}. 

LINX instead allows one to perform the fully marginalized BBN-only analysis with the PArthENoPE rates (solid red lines in Fig.~\ref{fig:parameter_evolution}).  The corresponding LINX result for constant $\sigma_{\rm th}$ is shown in pink and is in good agreement with the PArthENoPE results in Ref.~\cite{schoneberg2024}.  For both $\Omega_b h^2$ and $\Omega_b h^2 \texttt{+} N_\text{eff}$, proper marginalization of the nuclear rates increases the precision on  $\Omega_b h^2$ 
 by $\sim15$--$30\%$. 

\noindent
\noindent\textbf{Joint CMB\texttt{+}BBN Analyses.---}
LINX can also be used in conjunction with a CMB Boltzmann code to provide a joint inference of cosmological parameters.  We use the Boltzmann code CLASS and include one massive neutrino with mass $m_{\nu}=\SI{0.06}{eV}$, consistent with analyses in Ref.~\cite{Planck2018}.

A schematic of the joint likelihood used for a Bayesian inference of $\Omega_b h^2 \texttt{+} N_\text{eff}$ is included in Fig.~\ref{fig:likelihood}.
Given $\Omega_b h^2$, $N_\text{eff}$, and BBN nuisance parameters, LINX computes D/H$^\text{pred}$ and $\text{Y}_\text{P}^\text{pred}$ to evaluate $\mathcal{L}_\text{BBN}$ in Eq.~\eqref{eq:BBN_like}. 
The same $\Omega_b h^2$, $N_{\rm{eff}}$, and $\text{Y}_\text{P}^\text{pred}$---with the other $\Lambda$CDM 
and Planck nuisance parameters---are passed to CLASS and the Planck likelihood to predict power spectra and obtain $\mathcal{L}_\text{CMB}$. 
The joint likelihood is then used for Bayesian inference.
Importantly, $\text{Y}_\text{P}^\text{pred}$needs to be passed to CLASS from a BBN solver, since it has a significant impact on the predicted power spectrum.

\begin{figure*}[!t]
    \centering
    \includegraphics[width=0.32\textwidth]{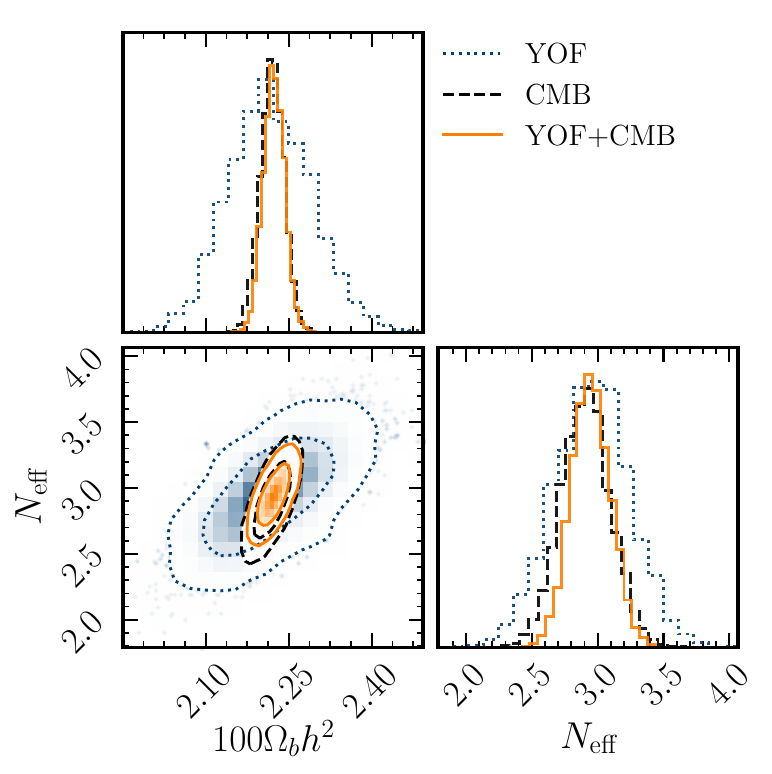}
    \includegraphics[width=0.332\textwidth]{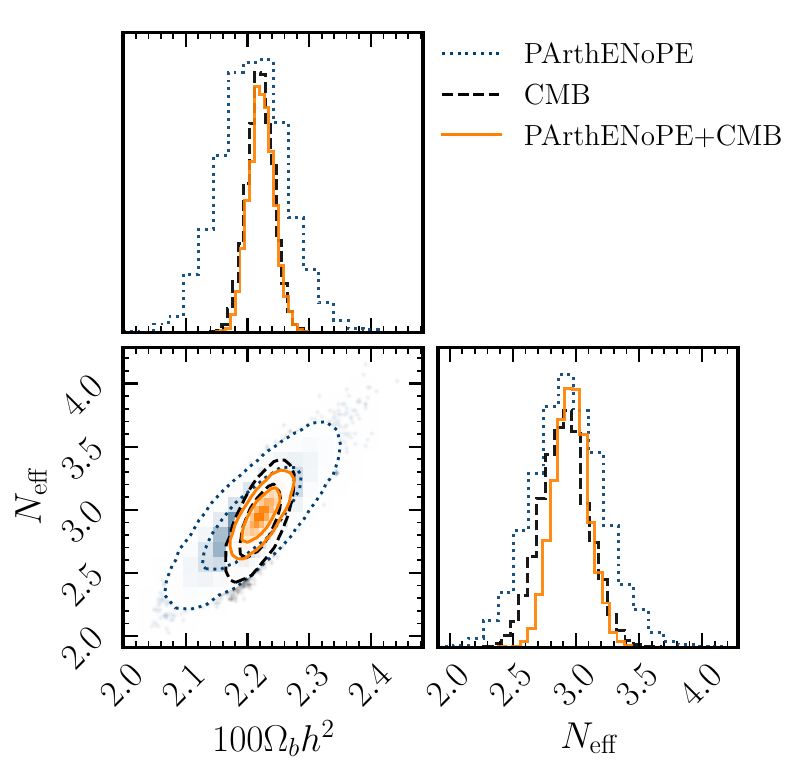}
    \includegraphics[width=0.32\textwidth]{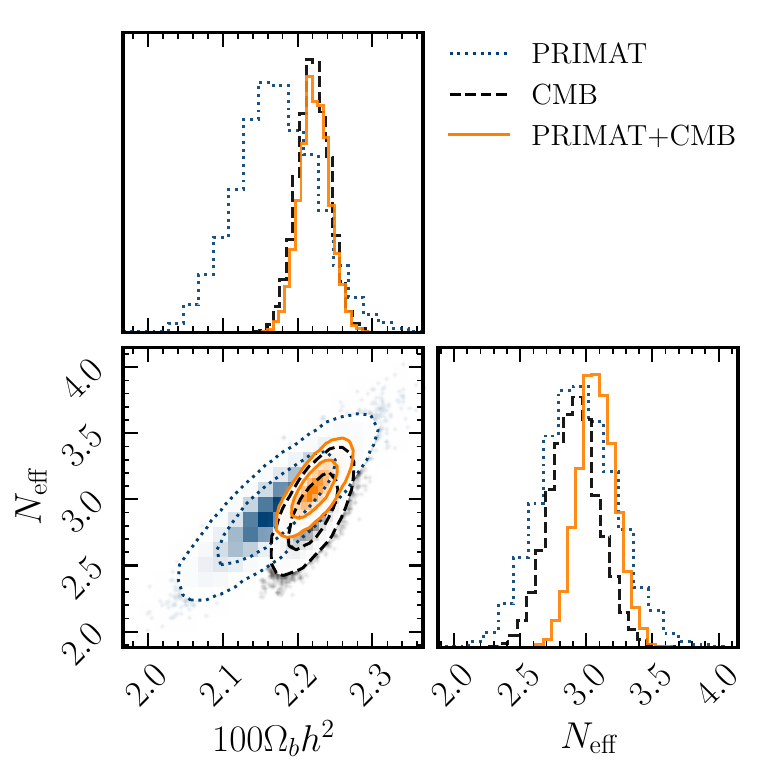}
    \caption{68\% and 95\% contours in the $\Omega_bh^2-N_{\rm{eff}}$ plane for YOF~(left), PArthENoPE~(center), and PRIMAT~(right) nuclear reaction networks.  In each panel, the light-blue dotted contours correspond to the BBN-only analysis, the black-dashed to the CMB-only analysis, and the orange-solid to the joint CMB\texttt{+}BBN analysis. 
 All CMB and BBN nuisance parameters are marginalized over. When using the PRIMAT network, the resulting value of $N_{\rm{eff}}$ in the joint analysis is pushed high.  When using the YOF network, the constraining power of BBN is minimal.}
    \label{fig:Neff}
\end{figure*}

Throughout, we use the \texttt{Plik} TT+TE+EE likelihood \cite{Planck_2018_likelihoods}, which contains high-$\ell$ multipoles, and the \texttt{commander} lowT and \texttt{simall} lowE likelihoods at low-$\ell$, for the CMB likelihood.  
In our joint analyses, we sample 40--41 parameters in the joint analysis, depending on whether $N_{\rm{eff}}$ is fixed and use \texttt{Plik}-recommended priors for the CMB nuisance parameters (or priors from Ref.~\cite{Planck2013} if unavailable). Broad, flat priors are used for the other CMB model parameters. 
Our inferred $100\Omega_b h^2$ and $\Omega_b h^2$\texttt{+}$N_\text{eff}$ with only Planck data are included in Fig.~\ref{fig:parameter_evolution} in the ``CMB-only" column for reference and are in excellent agreement with Ref.~\cite{Planck2018}.

The joint analysis results are shown under the ``CMB\texttt{+}BBN'' column in Fig.~\ref{fig:parameter_evolution}. The square markers show the inferred $100\Omega_b h^2$, holding $N_{\rm{eff}}$ fixed.  
For all three reaction networks, there is a shift in the medians and error bars towards the CMB-preferred value of $100\Omega_b h^2 = 2.231^{+0.015}_{-0.014}$.  Compared to the BBN-only results, the error bars are reduced in the joint results; this is most pronounced for the joint constraint with the YOF network, where the error bars shrink by a factor of $\sim3$.  
The PRIMAT joint result ($2.218^{+0.012}_{-0.012}$) is most discrepant with the CMB-only result, due to the low BBN-only preferred value for $100\Omega_b h^2$ ($2.191^{+0.022}_{-0.021}$). 

A simplified version of this analysis is performed in Ref.~\cite{Pitrou_2018} using the PRIMAT network.  For that study, the Planck posterior \cite{Planck2018} for $\Omega_b h^2$ is used as a prior in the BBN-only likelihood, so only the BBN likelihood must be computed for each sample.  When using LINX to perform this simplified analysis using the Planck 2018 result of $100\Omega_b h^2 = 2.236 ^{+0.015}_{-0.015}$ \cite{Planck2018} as the prior, we find that $100\Omega_b h^2 = 2.215^{+0.011}_{-0.011}$, which is slightly lower than the results with a full CMB analysis.

Next, we vary $N_{\rm{eff}}$ in addition to the six $\Lambda$CDM parameters, the BBN rate uncertainties, and the CMB nuisance parameters. 
For the PArthENoPE and YOF rates, including a BBN likelihood in this analysis does not dramatically shift many of the cosmological parameters away from their inferred values in a CMB-only analysis.  
This is because of the good agreement between the PArthENoPE network's predictions and measurements, as well as YOF's wide error bars on the reactions in its network, which decreases its constraining power. 
These effects are illustrated in the first two panels 
of Fig.~\ref{fig:Neff}, where the 
individual posteriors overlap with each other.    
These results are also included in Fig.~\ref{fig:parameter_evolution} in the ``CMB\texttt{+}BBN" column for comparison.

However, the individual posteriors for the PRIMAT network, shown in the third panel of Fig.~\ref{fig:Neff}, are in slight discrepancy with each other, which influences the inferred values of $\Omega_b h^2$ and $N_{\rm eff}$. The  inferred value for $N_{\rm{eff}}$ increases 
to $3.08^{+0.15}_{-0.14}$, from the CMB-only prediction of $2.95^{+0.19}_{-0.18}$ and the BBN-only prediction of $2.93^{+0.30}_{-0.27}$.  
The correlation between $N_{\rm{eff}}$ and $\Omega_b h^2$, and the resulting adjustment of the inferred values away from those attained in a CMB-only analysis, are responsible for the higher inferred value of $N_{\rm{eff}}$.  

\begin{figure}[t]
    \centering
    \includegraphics[width=\linewidth]{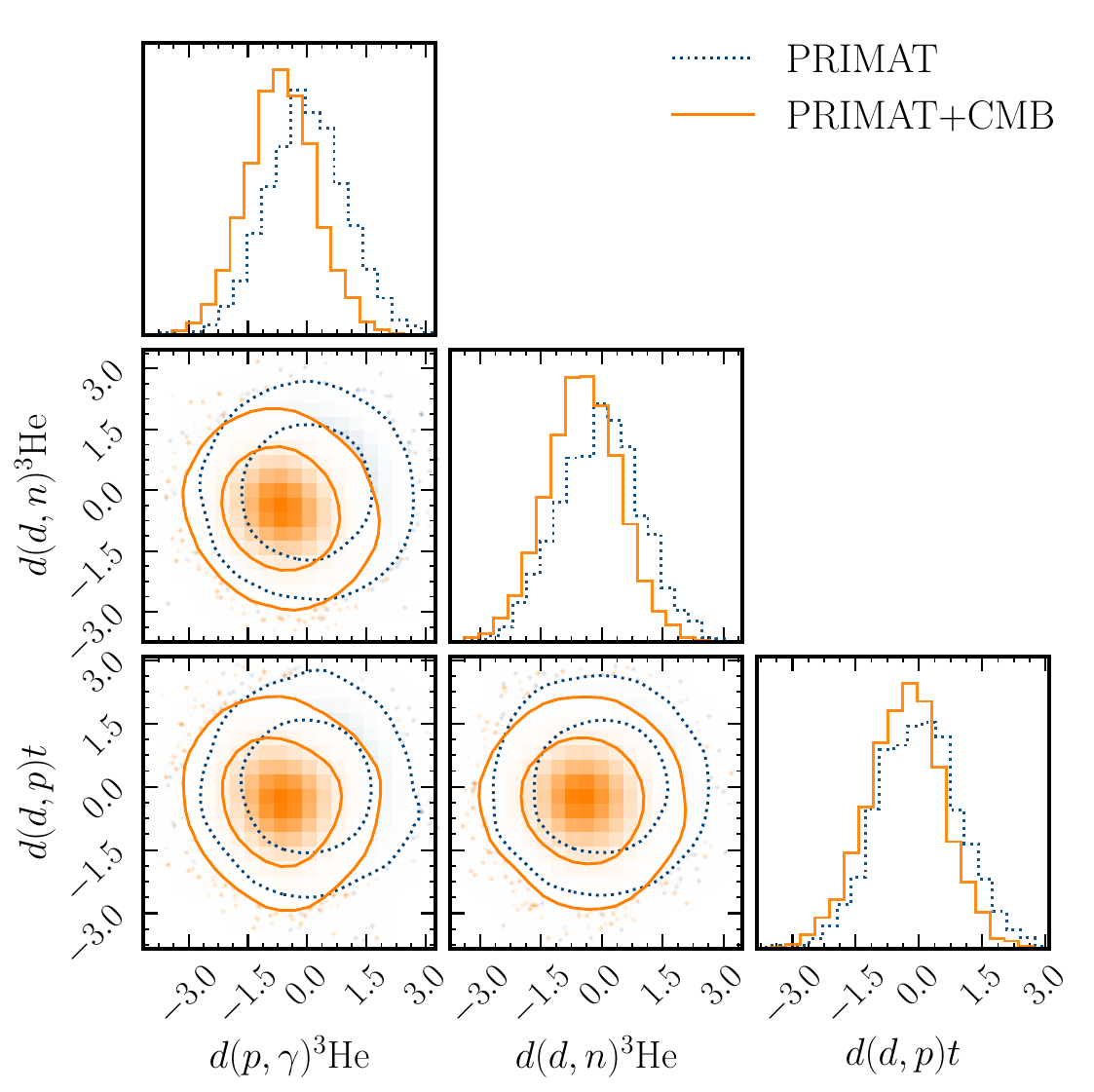}
    \caption{68\% and 95\% posterior contours for the nuisance parameters $q_i$ that determine the scaling of three key deuterium burning rates in the PRIMAT network.  The BBN-only analysis (blue, dashed) is compared to a joint CMB\texttt{+}BBN analysis (orange, solid), allowing $\Omega_b h^2$ to float and fixing $N_{\rm{eff}}= 3.045$. In the joint analysis, these rates are pushed away from their central values to attain a better joint fit with the CMB\@.}
    \label{fig:deuterium_rates}
\end{figure}

Ref.~\cite{Planck2018} observed a similar effect on the best-fit parameters when using the PRIMAT rates and a constant theory uncertainty $\sigma_{\rm th}$ on D/H and Y$_\mathrm{P}$.   
However, a fully marginalized analysis with LINX allows one to uncover correlations between nuisance parameters, including how nuclear rates have to adjust to accommodate the preferred region for the CMB\texttt{+}BBN fit. 
Fig.~\ref{fig:deuterium_rates} shows the marginalized posterior for three deuterium burning reactions ($d(p,\gamma)$$^3\mathrm{He}$, $d(d,n)$$^3\mathrm{He}$, and $d(d,p)t$) for the $\Omega_b h^2$ analysis. 
Decreased rates for all three reactions are required to bring BBN into better agreement with Planck data using PRIMAT\@. 

The joint analyses discussed above also have implications for the $\Lambda$CDM\texttt{+}$N_{\rm{eff}}$ parameters that do not have direct impacts on BBN\@.  When $N_{\rm{eff}}$ is allowed to float, including BBN in a joint analysis leads to smaller error bars in $N_{\rm{eff}}$, which can propagate to noticeably smaller error bars in these parameters.  For instance, when $N_{\rm{eff}}$ floats, the joint analysis gives error bars on the Hubble parameter and dark matter density that are $\sim 30\%$ smaller than the error bars on their CMB-only determinations.

\noindent
\noindent\textbf{Conclusions.---}This paper achieves four important goals:
\begin{enumerate}
    \item For the first time, we are able to perform cosmological parameter estimation using the CMB and BBN with a truly joint likelihood.  These are the most consistent estimates obtained to date.  This is our primary result.
    \item We demonstrate that there is much to be learned from proper treatment of nuisance parameters in a joint analysis such as this.  For example, in Fig.~\ref{fig:deuterium_rates}, we illustrate how the nuisance parameters adjust in response to the mild deuterium tension present in the PRIMAT network, providing physical insights into the tension and perhaps even directions for future work in better understanding this tension.
    \item We explore, in a self-consistent manner, systematics that arise from using different reaction networks.  This is different than the approach in Ref.~\cite{schoneberg2024}, which was subject to limitations of existing BBN codes so that direct comparisons could not be made between the PRIMAT and NACREII networks, whose rates could be marginalized, and the PArthENoPE network, whose rates were more difficult to marginalize. 
    \item {We demonstrate that marginalizing over the nuclear reaction rates is essential for proper estimation of error bars in BBN-only analyses.  Previous analyses have used an estimation scheme that was too conservative for these analyses, and had no way of knowing their estimates were too conservative, until we performed this marginalization.  We note that the variation in $\sigma_\text{D/H}$ is unknown for any model other than $\Lambda$CDM\texttt{+}$N_\text{eff}$, either as a function of these parameters or as a function of parameters in a new-physics model.  Since the only way to obtain a reliable estimate for $\sigma_{\text{D/H}}$ is to perform the sampling described above in the first place, there is no practical reason not to perform a proper marginalization over BBN nuisance parameters using LINX, \textit{e.g.}\ within the context of an MCMC sampler. }
\end{enumerate}

This study reveals the importance of proper accounting of the CMB and BBN nuisance parameters in a joint CMB\texttt{+}BBN analysis, as well as including the impacts of BBN on the determination of $\Omega_b h^2$ and Y$_{\rm{P}}$.  Consistent analyses like those described in this manuscript are essential for testing cosmological models. The CMB\texttt{+}BBN results come at little computational cost; LINX is fast enough that the time taken to evaluate the joint likelihood is dominated by the Boltzmann solver.  We include a complete index of the standard cosmological results in Appendix~\ref{app:results}.  Our analyses also explore systematic effects from choosing different reaction networks---depending on which network is used, BBN can have little constraining power, can validate CMB analyses, or can pull parameters away from their CMB-only values.  

Moving forward, LINX can be used in conjunction with cosmology sampling frameworks such as cosmomc~\cite{Lewis_2002}, MontePython~\cite{brinckmann2018montepython}, and other tools for cosmology data analysis.  
Additionally, LINX greatly facilitates the extension of joint CMB\texttt{+}BBN fits to new-physics scenarios, including electromagnetically coupled WIMPs~\cite{Sabti:2019mhn,Sabti:2021reh,Giovanetti:2021izc,An:2022sva}, neutrino-coupled WIMPs~\cite{Berlin:2017ftj,Giovanetti:2024orj}, \SI{}{\mega\eV}-scale axion-like particles~\cite{Depta:2020zbh}, and hidden sector particles such as millicharged particles with dark radiation~\cite{Adshead:2022ovo}. 
LINX can also be used to obtain an improved joint BBN\texttt{+}BAO determination of the Hubble constant, a crucial piece of the Hubble tension puzzle~\cite{Schoneberg:2019wmt}, or to explore other upcoming and existing measurements of various primordial abundances in extended likelihoods (\textit{e.g.}\ $^7$Li, $^3$He, or even other measured values of Y$_{\rm{P}}$ as in Refs.~\cite{Aver_2021_LeoP,Burns_2023_lepton,Escudero_2023}). Finally, we note the value of this work in light of upcoming CMB experiments like the Simons Observatory~\cite{simons} and CMB-S4~\cite{CMB-S4}.   Given the increasing number of nuisance parameters expected for these experiments, having an efficient pipeline for joint CMB and BBN analyses is essential, even in $\Lambda$CDM.  The methodology presented here can be straightforwardly extended to analyses for these new upcoming data. Further, while both experiments will improve upon the determination of $N_{\rm{eff}}$, $N_{\rm{eff}}$ as determined by BBN contains different information than $N_{\rm{eff}}$ as determined from the CMB~\cite{abazajian2016cmbs4sciencebookedition}, and so the foundation laid in this work will remain relevant as these new experiments begin taking data.

\vspace{0.2in}
\section*{Acknowledgements}
We thank the PRyMordial team (Anne-Katherine Burns, Tim Tait, and Mauro Valli) for early access to their code and assistance with setup and usage throughout the beta-testing phase.  We thank Nick dePorzio, Colin Hill, and Zilu Zhou for help with CLASS\@.  We further thank Zilu Zhou for help with the \texttt{Plik} likelihoods.  We thank Mark Paris for information about the measurement of deuterium reaction rates.  We thank the JAX team and the individuals who assisted us with JAX and JAX packages, including Dan Foreman-Mackey, Peter Hawkins, and Patrick Kidger.  
We thank Julien Lesgourgues and Nils Sch\"{o}neberg for comments on a draft of this paper.  
ML is supported by the Department of Energy~(DOE) under Award Number DE-SC0007968 as well as the Simons Investigator in Physics Award. 
HL was supported by the Kavli Institute for Cosmological Physics and the University of Chicago through an endowment from the Kavli Foundation and its founder Fred Kavli, and Fermilab operated by the Fermi Research Alliance, LLC under contract DE-AC02-07CH11359 with the U.S. Department of Energy, Office of Science, Office of High-Energy Physics.  
SM is partly supported by the U.S. Department of Energy, Office of Science, Office of High Energy Physics of U.S. Department of Energy under grant Contract Number  DE-SC0012567. 
JTR is supported by NSF grant PHY-2210498.
This material is based upon work supported by the NSF Graduate Research Fellowship under Grant No.~DGE1839302. 
This work is supported by the National Science Foundation under Cooperative Agreement PHY-2019786 (The NSF AI Institute for Artificial Intelligence and Fundamental Interactions, \href{http://iaifi.org/}{http://iaifi.org/}).
This research was supported in part by Perimeter Institute for Theoretical Physics. Research at Perimeter Institute is supported by the Government of Canada through the Department of Innovation, Science and Economic Development and by the Province of Ontario through the Ministry of Research, Innovation and Science. 
This work was performed in part at the Aspen Center for Physics, which is supported by NSF grants PHY-1607611 and PHY-2210452.
This research was supported in part by grant NSF PHY-2309135 to the Kavli Institute for Theoretical Physics~(KITP).  
The work presented in this paper was performed on computational resources managed and supported by Princeton Research Computing.  This work was supported in part through the NYU IT High Performance Computing resources, services, and staff expertise.  This work makes use of the corner~\cite{corner}, diffrax~\cite{kidger2021on}, dynesty~\cite{Speagle_2020,speagle_2023}, equinox~\cite{kidger2021equinox}, JAX~\cite{jax2018github,deepmind2020jax}, matplotlib~\cite{Hunter:2007}, numpy~\cite{harris2020array}, schwimmbad~\cite{schwimmbad}, and scipy~\cite{2020SciPy-NMeth} Python packages.
\bibliography{references}

\clearpage 
\onecolumngrid
\appendix

\section{Computational Analysis Details}\label{app:details}
We use nested sampling~\cite{Skilling_2004,Skilling_2006}, as implemented by dynesty~\cite{Speagle_2020,speagle_2023}, with  a static sampling method~\cite{Feroz_2009,Neal_2003,Handley_2015a,Handley_2015b}.  In the analyses reported in this paper and in Appendix~\ref{app:results}, we sample using 1000~(500) live points for BBN-only~(CMB\texttt{+}BBN) analyses and stop when the change in the log of the remaining evidence is $0.5$.

The BBN-only analyses take place on 36 Lenovo SD650 standard-memory cores.  Including marginalization, these scans take roughly 3-6 CPU-hours to complete (5-10 minutes wall-clock time).  The joint analyses are more resource-intensive due to CLASS's runtime, running on 192 Lenovo SD650 standard-memory cores in $\sim$8,500 CPU-hours ($\sim$2 days wall-clock time).

\section{Tabulated Results}\label{app:results}

\setcounter{equation}{0}
\setcounter{figure}{0}
\setcounter{table}{0}
\renewcommand{\theequation}{B\arabic{equation}}
\renewcommand{\thefigure}{B\arabic{figure}}
\renewcommand{\thetable}{B\arabic{table}}

Table~\ref{tab:full_results} compiles the results of all analyses performed in this work. The first block  includes results from BBN-only analyses without any marginalization of BBN nuisance parameters.  These are only intended to be compared with results from marginalizing nuisance parameters in the next block, as well as for comparison with existing results in the literature.  \textit{Parameters in this block should not be taken as results or used in future work}---only the results that include marginalization over the nuclear rates should be used.

Similarly, we include results from CMB-only analyses (with Planck nuisance parameters marginalized over) for easy comparison with the results from the CMB\texttt{+}BBN joint analyses (with both Planck and BBN nuisance parameters marginalized over) in the last two blocks. 
We also show results from other studies in the literature when they can be directly compared with LINX results. 
These come from Ref.~\cite{schoneberg2024} for BBN-only results and Ref.~\cite{Planck2018} for CMB-only results. 
In all cases where a comparison is possible, the results from our analyses are in excellent agreement with existing results. 

\begin{table}[h]
    \centering
    \renewcommand{\arraystretch}{1.5} 
    \begin{tabular}{ccccccccc}
        \multicolumn{1}{p{2cm}}{\centering BBN Rates} & \multicolumn{1}{p{2.5cm}}{\centering $\boldsymbol{\nu}_\text{BBN}$ Marginalization} & \multicolumn{1}{p{2cm}}{\centering Planck} & \multicolumn{1}{p{2.5cm}}{\centering $100\Omega_b h^2$} & \multicolumn{1}{p{2.5cm}}{\centering $100\Omega_b h^2$ Reference} & \multicolumn{1}{c}{$N_\text{eff}$} & \multicolumn{1}{p{1.5cm}}{\centering $N_\text{eff}$ Reference} & \multicolumn{1}{p{1.5cm}}{\centering Reference}& 
         \\

        \hline
        \multicolumn{9}{c}{\centering \textbf{BBN only, no marginalization over nuclear rates}} \\
        PRIMAT & \textcolor{red}{\faTimesCircle} & \textcolor{red}{\faTimesCircle} & $2.192 ^{+0.016}_{-0.016}$ & -- & \textcolor{red}{\faTimesCircle} & \textcolor{red}{\faTimesCircle} & -- & 
        \\
        PRIMAT & \textcolor{red}{\faTimesCircle} & \textcolor{red}{\faTimesCircle} & $2.173^{+0.051}_{-0.050}$ & -- & $2.94^{+0.28}_{-0.27}$ & -- & -- & 
        \\
        YOF & \textcolor{red}{\faTimesCircle} & \textcolor{red}{\faTimesCircle} & $2.232^{+0.017}_{-0.016}$ & $2.234^{+0.016}_{-0.016}$ & \textcolor{red}{\faTimesCircle} & \textcolor{red}{\faTimesCircle} & \cite{schoneberg2024} & 
        \\
        YOF & \textcolor{red}{\faTimesCircle} & \textcolor{red}{\faTimesCircle} & $2.209 ^{+0.056}_{-0.050}$ & -- & $2.91^{+0.32}_{-0.26}$ & -- & -- & 
        \\
        PArthENoPE & \textcolor{red}{\faTimesCircle} & \textcolor{red}{\faTimesCircle}  & $2.229^{+0.016}_{-0.015}$ & -- & \textcolor{red}{\faTimesCircle} & \textcolor{red}{\faTimesCircle} & -- & 
        \\
        PArthENoPE & \textcolor{red}{\faTimesCircle} & \textcolor{red}{\faTimesCircle} & $2.209^{+0.054}_{-0.052}$ & -- & $2.93^{+0.29}_{-0.28}$ & -- & -- & 
        \\
        \hline

        \multicolumn{9}{c}{\centering \textbf{BBN only, marginalized over nuclear rates}} \\
        PRIMAT & \textcolor{green}{\faCheckCircle}  & \textcolor{red}{\faTimesCircle} & $2.191^{+0.022}_{-0.021}$ & $2.195^{+0.021}_{-0.021}$ & \textcolor{red}{\faTimesCircle} & \textcolor{red}{\faTimesCircle} & \cite{schoneberg2024} & 
        \\
        PRIMAT & \textcolor{green}{\faCheckCircle} & \textcolor{red}{\faTimesCircle}  & $2.171^{+0.056}_{-0.051}$ &$2.172^{+0.055}_{-0.055}$ & $2.93^{+0.30}_{-0.27}$ & $2.92^{+0.28}_{-0.28}$ & \cite{schoneberg2024} & 
        \\
        YOF & \textcolor{green}{\faCheckCircle} & \textcolor{red}{\faTimesCircle}  & $2.233^{+0.060}_{-0.056}$ & $2.231^{+0.055}_{-0.055}$ & \textcolor{red}{\faTimesCircle} & \textcolor{red}{\faTimesCircle} & \cite{schoneberg2024} & 
        \\
        YOF & \textcolor{green}{\faCheckCircle} & \textcolor{red}{\faTimesCircle} & $2.214^{+0.079}_{-0.073} $ & $2.212^{+0.072}_{-0.072}$ & $2.96^{+0.28}_{-0.30}$ & $2.93^{+0.27}_{-0.27}$ & \cite{schoneberg2024} & 
        \\
        PArthENoPE & \textcolor{green}{\faCheckCircle} & \textcolor{red}{\faTimesCircle} & $2.228^{+0.024}_{-0.023}$ & -- & \textcolor{red}{\faTimesCircle} & \textcolor{red}{\faTimesCircle} & -- & 
        \\
        PArthENoPE & \textcolor{green}{\faCheckCircle} & \textcolor{red}{\faTimesCircle} & $2.209 ^{+0.053}_{-0.053}$ & -- & $2.93_{-0.28}^{+0.28}$ & -- & -- &
        \\

        \hline
        \multicolumn{9}{c}{\centering \textbf{CMB only, marginalized over Planck nuisance parameters}} \\
        \textcolor{red}{\faTimesCircle} & \textcolor{red}{\faTimesCircle} & \textcolor{green}{\faCheckCircle} & $2.231_{-0.014}^{+0.015}$ & $2.236^{+0.015}_{-0.015}$ & \textcolor{red}{\faTimesCircle} & \textcolor{red}{\faTimesCircle} & \cite{Planck2018}& \\
        \textcolor{red}{\faTimesCircle} & \textcolor{red}{\faTimesCircle} & \textcolor{green}{\faCheckCircle} & $2.223 _{-0.021}^{+0.021}$ & $2.224^{+0.022}_{-0.022}$ & $2.95_{-0.18}^{+0.19}$ & $2.92^{+0.19}_{-0.19}$ & \cite{Planck2018} &\\

        \hline

        \multicolumn{9}{c}{\centering \textbf{CMB\texttt{+}BBN, marginalized over both nuclear rates and Planck nuisance parameters}} \\
        PRIMAT & \textcolor{green}{\faCheckCircle} & \textcolor{green}{\faCheckCircle} & $2.218 ^{+0.012}_{-0.012}$ & -- & \textcolor{red}{\faTimesCircle} & \textcolor{red}{\faTimesCircle} & -- &  \\
        PRIMAT  & \textcolor{green}{\faCheckCircle} & \textcolor{green}{\faCheckCircle} & $2.222 _{-0.019}^{+0.020}$ & -- & $3.08_{-0.14}^{+0.15}$ & -- & -- &  \\
        YOF  & \textcolor{green}{\faCheckCircle} & \textcolor{green}{\faCheckCircle} & $2.234_{-0.014}^{+0.014}$ & --  & \textcolor{red}{\faTimesCircle} & \textcolor{red}{\faTimesCircle} & -- &  \\
        YOF & \textcolor{green}{\faCheckCircle} & \textcolor{green}{\faCheckCircle} & $2.223^{+0.019}_{-0.019}$ & --  & $2.94 _{-0.15}^{+0.16}$ & -- & -- &  \\
        PArthENoPE  & \textcolor{green}{\faCheckCircle} & \textcolor{green}{\faCheckCircle} & $2.232_{-0.012}^{+0.013}$ & -- & \textcolor{red}{\faTimesCircle} & \textcolor{red}{\faTimesCircle} & -- &  \\
        PArthENoPE & \textcolor{green}{\faCheckCircle} & \textcolor{green}{\faCheckCircle} & $2.222_{-0.020}^{+0.020}$ & -- & $2.96_{-0.14}^{+0.13}$ & -- & -- &  \\
        \hline
    \end{tabular}
    \caption{Median inferred parameters and 68\% credible intervals for:~\emph{1)} BBN-only with no marginalization over BBN nuisance parameters, \emph{2)}~BBN-only with marginalization over nuisance parameters, \emph{3)}~CMB-only with marginalization over Planck nuisance parameters, and \emph{4)}~BBN\texttt{+}CMB joint analyses with marginalization over both Planck and BBN nuisance parameters.  The symbol \textcolor{red}{\faTimesCircle} indicates that a given item is not included in the analysis, while the symbol \textcolor{green}{\faCheckCircle} indicates that item is included in the analysis. `--' indicates that a reference for comparison does not exist---references are provided only where equivalent analyses have been performed, rather than simplified analyses.  We use D/H from Ref.~\cite{Cooke_2018} and Y$_{\rm{P}}$ from Ref.~\cite{Aver_2015} as measured values for the BBN likelihood and also in the reference results. Parameters from Ref.~\cite{Planck2018} are taken for TTTEEE+lowT+lowE\@.}
    \label{tab:full_results}
\end{table}

\end{document}